\documentclass[usenatbib]{mn2e} \usepackage{epsfig}
\newcommand{\lsim}{\lower 2pt \hbox{$\, \buildrel {\scriptstyle
<}\over {\scriptstyle \sim}\,$}}  \newcommand{\gsim}{\lower 2pt
\hbox{$\, \buildrel {\scriptstyle >}\over {\scriptstyle \sim}\,$}}

\bibpunct{(}{)}{;}{a}{,}{,} \title[Optical spectroscopy of Arp220: the
star formation history of the closest ULIRG.]{Optical spectroscopy of
Arp220: the star formation history of the closest ULIRG}
\author[J.Rodr\'iguez Zaur\'in, C.N.Tadhunter and R.M.Gonz\'alez
Delgado$^{2}$]{J.Rodr\'iguez
Zaur\'in$^{1}$\thanks{E-mail:jrzaurin@sheffield.ac.uk}, C.N
Tadhunter$^{1}$ and R.M.Gonz\'alez Delgado$^{2}$\\ $^{1}$Department of
physics and Astronomy, University of Sheffield, Sheffield S3 7RH\\
$^{2}$Instituto de Astrofisica de Andalucia(CSIC), P.O.Box 3004, 18080
Granada, Spain}
\begin{document}

\pagerange{\pageref{firstpage}--\pageref{lastpage}} \pubyear{2002}

\maketitle

\label{firstpage}

\begin{abstract}
We present long-slit, optical spectra of the merging system Arp 220,
taken with the William Herschel Telescope (WHT) on La Palma. These
data were taken as a part of a large study of ultraluminous infrared
galaxies (ULIRGs) with the aim of investigating the evolution and star
formation histories of such objects. Spectral synthesis modelling has
been used to estimate the ages of the stellar populations found in the
diffuse light sampled by the spectra. As the closest ULIRG, it proved
possible to perform a detailed study of the stellar populations over
the entire body of the object. The data show a remarkable uniformity
in the stellar populations across the full 65 arcsec covered by our
slit positions, sampling the measurable extent of the galaxy. The
results are consistent with a dominant intermediate-age stellar
population (ISP) with age 0.5 $<$ t$_{ISP}$ $\leq$ 0.9 Gyr that is
present at all locations, with varying contributions from a young
($\leq$ 0.1 Gyr) stellar population (YSP) component. However, it is
notable that while the flux contribution of the YSP component in the
extended regions is relatively small ($\leq$ 40\%), adequate fits in
the nuclear region are only found for combinations with a significant
contribution of a YSP component (22 - 63\% ). Moreover, while a low
intrinsic reddening ($E(B - V)$$\lsim$ 0.3) is found for the ISPs in
the extended regions, intrinsic reddening values as high as $E(B - V)
\sim $ 1.0 are required in the galactic center. This clearly reflects
the presence of a reddening gradient, with higher concentrations of
gas and dust towards the nuclear regions, coinciding with dust lanes
in the HST images. Overall, our results are consistent with models
that predict an epoch of enhanced star formation coinciding with the
first pass of the merging nuclei (represented by the ISP), with a
further episode of star formation occurring as the nuclei finally
merge together (represented by the YSP and ULIRG).


\end{abstract}


\section{Introduction}
Since the early 1980s, several studies have revealed the presence of
galaxies with the spectral energy distribution dominated by infrared
emission
\citep{Houck84,Houck85,Soifer84a,Soifer84b,Soifer87,Lefloch05}. These
galaxies are classified as Luminous (L$_{ir} > 10^{11}$ L$_{\odot}$)
or Ultraluminous ($L_{ir} > 10^{12}$ L$_{\odot}$) infrared galaxies
(LIRGs/ULIRGs). Their prodigious infrared emission is generally
attributed to the optical/UV light of luminous central sources
reprocessed by dust. Starburst activity is ongoing in most, if not
all, of these sources, coexisting with AGN activity in some cases
\citep{Genzel98,Surace99,Veilleux99}. Therefore these objects provide
us with an excellent oportunity to study both AGN and starburst
phenomena. Moreover ULIRGs are almost unvariably associated with
galaxy mergers and interactions (see \cite{Sanders96} for a
revision). Thus, they also represent ideal objects to test models of
galaxy evolution via major galaxy mergers.  (e.g. Mihos \& Hernquist,
1996; Barnes \& Hernquist, 1996; Springel et al., 2005.)

In this context Arp220, as the closest ULIRG (z = 0.018), is a key
object for our understanding of star-formation in galaxy mergers.
\begin{figure*}
\begin{minipage}{170mm}
\begin{tabular}{cc}
\hspace*{-1cm}\psfig{file=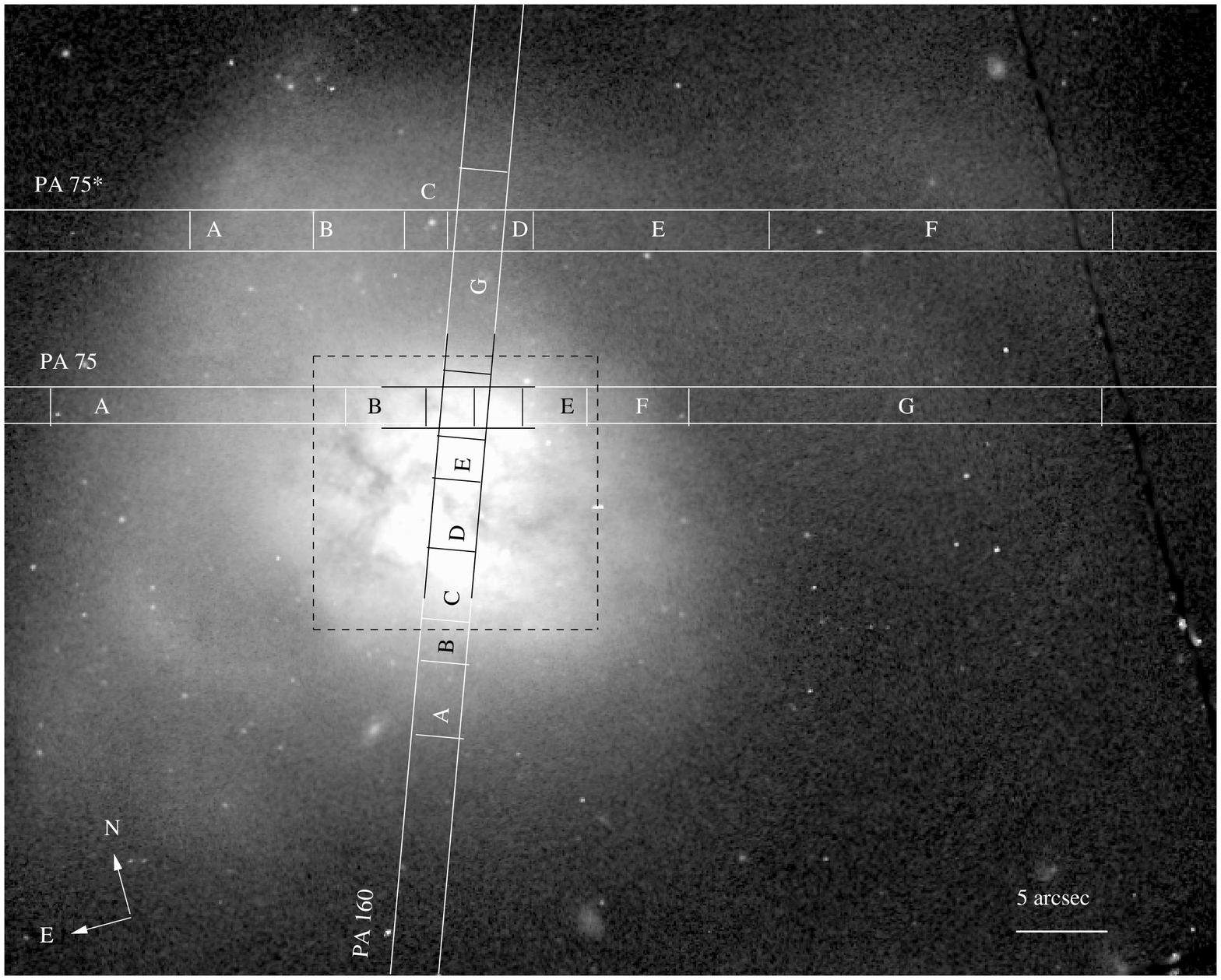,width=10cm,angle=0.}&
\psfig{file=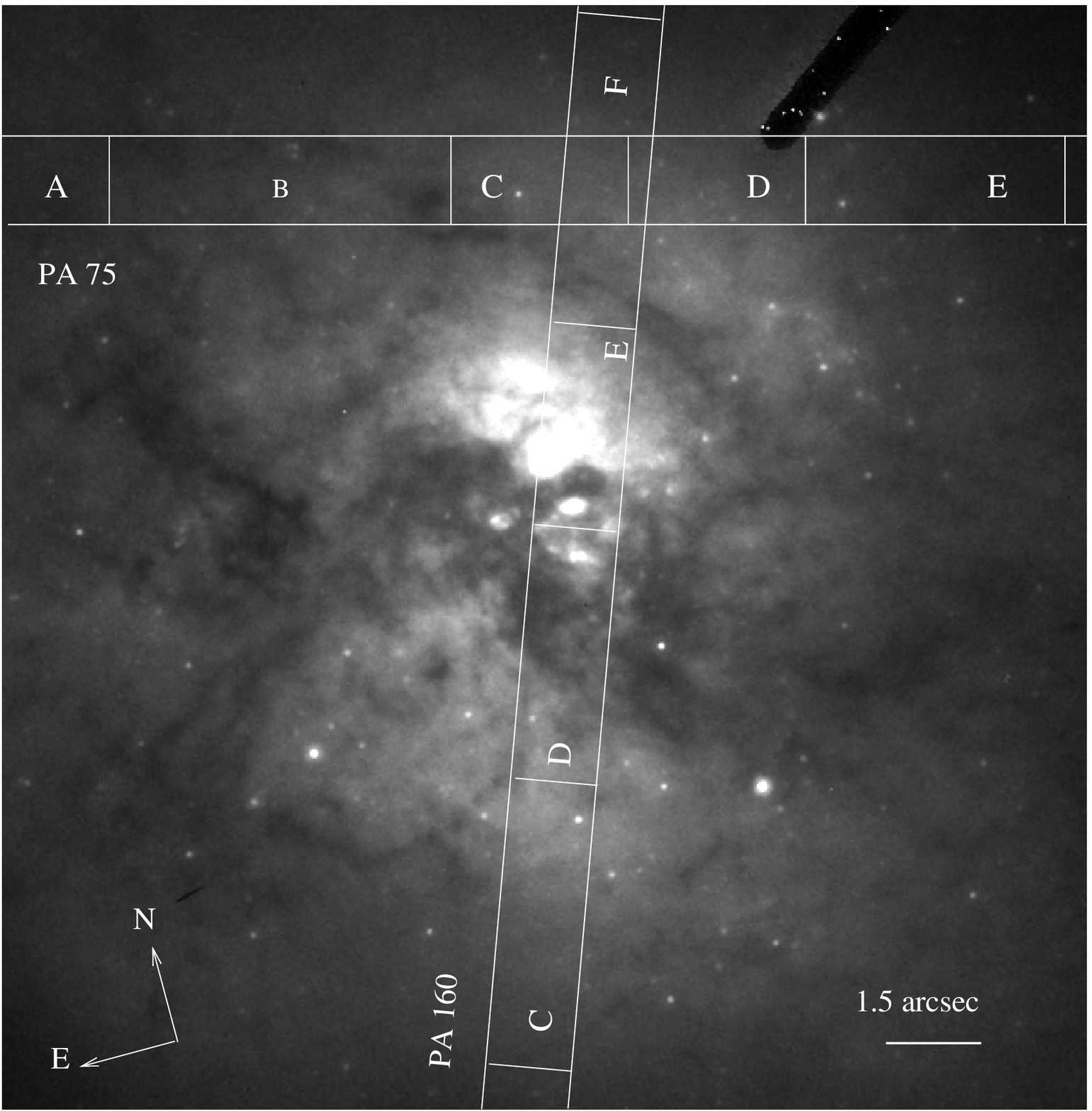,width=8cm,angle=0.}
\end{tabular}
\caption{Hubble space Telescope (HST) images taken with the F814W
filter, showing the locations of the slits and the extraction
apertures. Left: WFPC2 image showing both the nuclear region and the
extended regions. Right: ACS HRC Image showing the region within the
dashed-line box in the left pannel. }
\end{minipage}
\end{figure*} 
At infrared wavelengths, it appears as a double nucleus system
\citep{Graham90,Scoville98} separated 0.95 arcsec, with two tidal
tails visible at optical wavelengths \citep{Joseph85}. Although the
mid-IR spectra suggest starburst activity as the main source of power
for the infrared continuum \citep{Genzel98,Lutz99}, the galaxy is
classified as a LINER in the optical \citep{Veilleux99}. Arp 220 was
also observed by \cite{Scoville97} at 1.3 and 2.6 $\mu$m, sampling the
CO (2-1), CO (1-0) and dust continuum emission. They determine a total
H$_2$ mass of 3.2 $\times$ 10$^{10}$$M_{\odot}$, of which 2/3 is
contained within 250 pc and confined to a thin, centrally located,
dense disk, embedded in a CO disk extended by $\sim$1 Kpc . The
dynamics of the nuclear region are consistent with a collision between
prograde and retrograde disks overlapping with the main gas disk (
Mundel et al. 2001, see figure 7 in their paper). Recently,
\cite{Wilson06} performed a detailed photometric study of the star
clusters in Arp220. They found evidence that the cluster population
can be divided into two groups: a centrally located young population
of age t$_{YSP}$ $<$ $\lsim$ 10 Myr, and an intermediate-age
population of age $\sim$300 Myr spread towards the north of the
galaxy. However, given the photometric uncertainties, it is not
straightforward to break the age/reddening degeneracy problem
unambiguously in such studies, based only on colour-colour diagrams
with relatively few filters.

To date, there have been few spectroscopic studies of stellar
populations in ULIRGs. To remedy this situation we present in this
paper a detailed study of the stellar populations in Arp220, based on
spectroscopic observations of the extended diffuse light along three
slit positions. The results are discussed in the context of
evolutionary models of star formation in mergers.

Throughout the paper we assume $H_0= 71$ km s$^{-1}$, $\Omega_{0} =
0.27$ and $\Omega_{\Lambda} = 0.73$, resulting in a scale of 0.363 kpc
arcsec$^{-1}$ and a distance of 77.6 Mpc at z = 0.018.
\section{OBSERVATIONS AND DATA REDUCTION}
Spectra of Arp220 were taken in July 2005 and 2006 with the ISIS
dual-beam spectrograph on the 4.2-m William Herschel Telescope (WHT)
on La Palma. The R300B grating with the EEV12 CCD, and R316R grating
with MARCONI2 CCD were used on the blue and the red arms
respectively. A dichroic cutting at 5300\AA~was used during the
observations, leading to a useful wavelength range in the rest frame
of 3000 -- 5000\AA~ in the blue, and 5000 -- 7800\AA~ in the red. The
spectra were taken along three slits positions: PA 160, PA 75 and PA
75* (10 arsec offset to the north of PA 75), covering the nuclear
region and also the tidal tail observed towards the North-West of the
galaxy.  Details of the observations are given in Table 1. Figure 1
shows the slit positions superimposed on HST images taken with the
WFPC2 (HST proposal 6346, PI K.Borne) and the ACS HRC (HST proposal
9396, PI C.Wilson) camera. To find the precise locations of the slits
on the images we first convolved the images with Gaussian profiles to
simulate the seeing conditions during our WHT observations. Several
spatial profiles were then extracted from the images and compared with
those of spatial slices extracted from the spectra using a wavelength
range as close as posible to that of the images, until a match was
found. All exposures were taken with the slit aligned along the
parallactic angle, in order to minimize the effects of the
differential refraction.

The data were reduced (bias subtraction, flat field correction, cosmic
ray removal, wavelength calibration and flux calibration) and
straightened before the extraction of the individual spectra using the
standard packages in {\it IRAF} and the {\it STARLINK} packages {\it
FIGARO} and {\it DIPSO}. The wavelength calibration accuracy,
calculated as the mean value of the difference between the published
\citep{Osterbrock96} and the actual wavelengths of the night sky lines
was estimated as $<$ 0.34\AA~ and $<$ 0.45\AA~ in the blue and the red
respectively. The spectral resolution, calculated as the mean value of
the sky line width (FWHM) was 5.80\AA~ in the blue and 5.25\AA~ in the
red. The relative flux calibration accuracy, based on the observation
of several spectrophotometric standard stars, is estimated as $\pm$
5\% over the entire spectral range. This is confirmed by the excellent
match between the blue and the red spectra for all the apertures
extracted. Prior to the modelling, the spectra were corrected for
Galactic extinction using the \cite{Seaton79} reddening law and the
value of $E(B - V)$ = 0.051 from \cite{Schlegel98}.

\begin{table*}
\centering
\begin{minipage}{140mm}
\begin{tabular}{@{}llcllllll@{}}
\hline   Date  &Arm & set up & Slit PA  & Slit width & Exposure Time & Seeing\\   
             & & (CCD + gratting) &      &(arcsec)   & (sec) & (arcsec)\\    
\hline   
July 2005  & B & EEV12 + R300B & 160 & 1.5 & 900 & 1.85\\    
July 2005  & R & MARCONI2 + R316R & 160 & 1.5 & 900 & 1.85\\    
July 2006  & B & EEV12 + R300B & 75 & 1.5 & 900 & 0.65\\
July 2006  & R & MARCONI2 + R316R & 75 & 1.5 & 900 & 0.65\\   
July 2006  & B & EEV12 + R300B & 75* & 1.5 & 1200 & 0.75\\    
July 2006  & R & MARCONI2 + R316R & 75* & 1.5 & 1200 & 0.75\\    
\hline  \hline
\end{tabular}
\caption{Summary of the spectroscopic observations of Arp220.}
\label{Summary of the observations}
\end{minipage}
\end{table*}
\begin{figure}
\hspace*{-0.5cm}\psfig{file=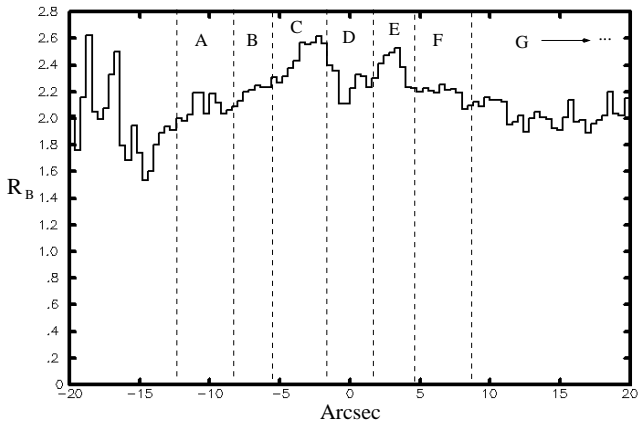,width=9cm,angle=0.}
\caption{The Balmer break ratio R$_{B}$ plotted as a function of
position along PA 160. The small variations of the ratio suggest a
remarkable uniformity for the stellar populations along the extension
covered by the slit position. The centroid of Ap D is used as the
reference point. Positive x = North, Negative x = South.}
\end{figure}
\begin{figure}
\psfig{file=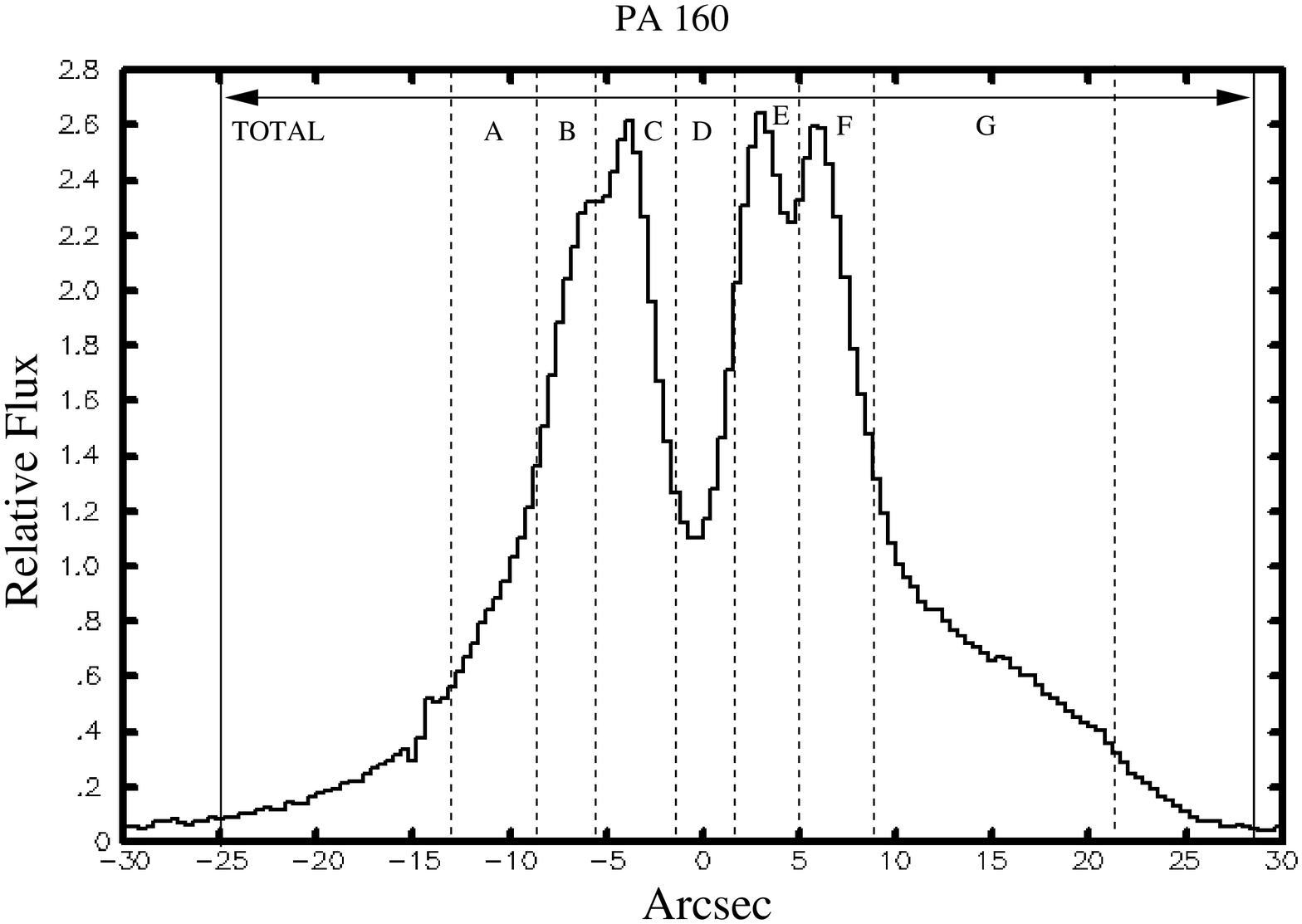,width=8.0cm,angle=0.}\\
\psfig{file=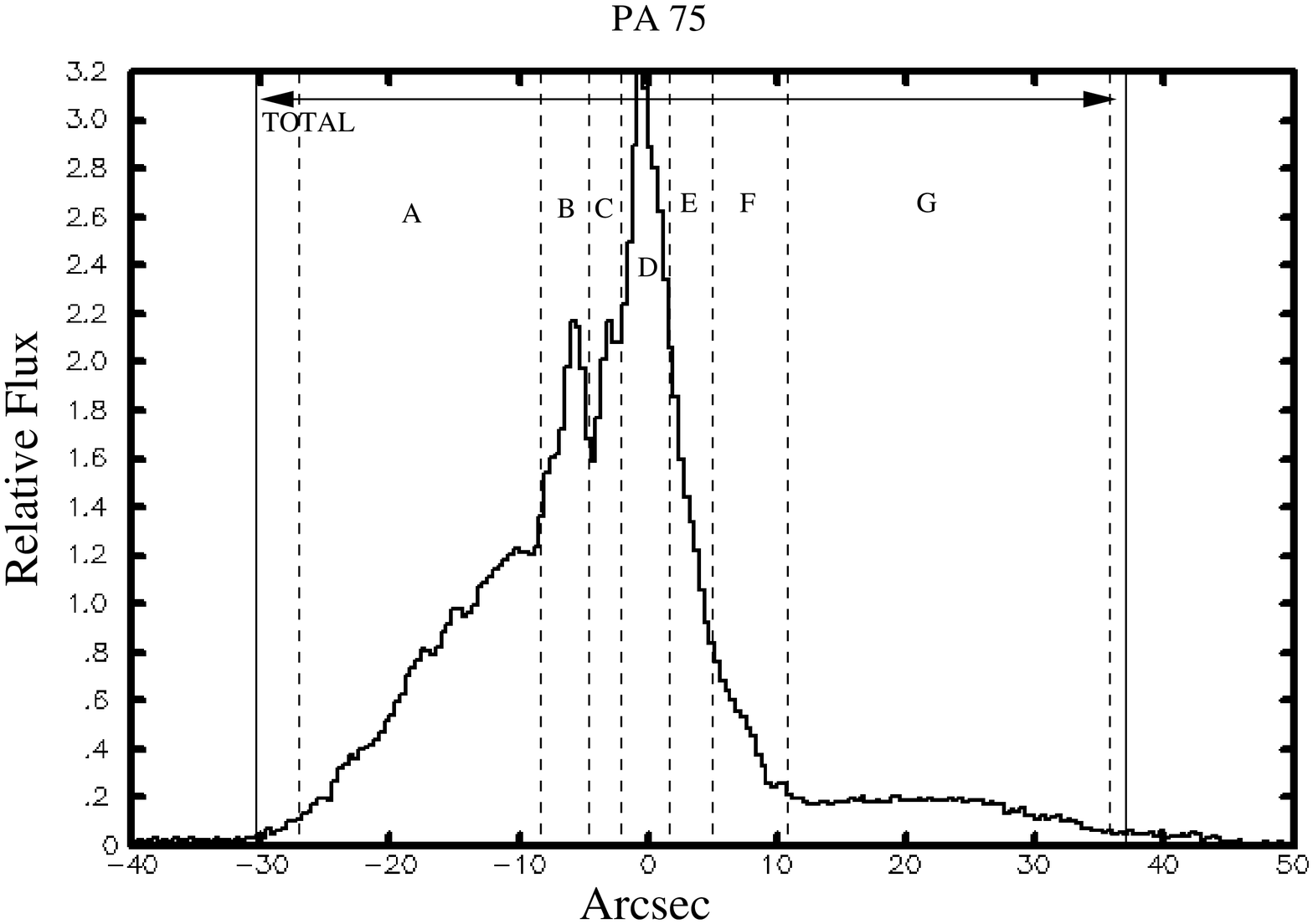,width=8.0cm,angle=0.}\\
\psfig{file=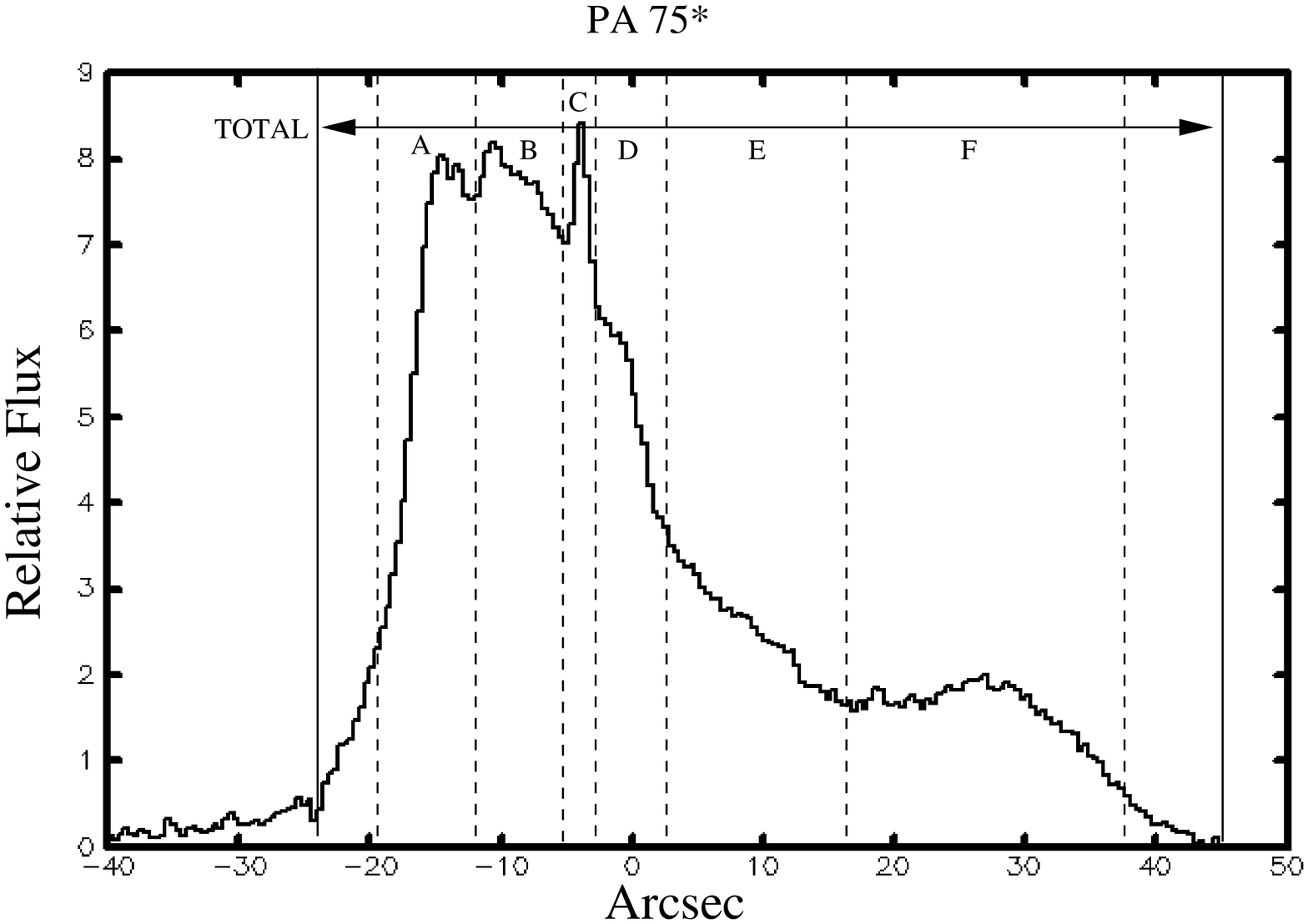,width=8.0cm,angle=0.}
\caption{Spatial profiles of the 2-D spectra for the wavelength range
4400 -- 4600\AA~showing the positions of the extraction apertures. The
centroid of Ap D is chosen as the reference point for the three slit
position, PA160, PA75 and PA75*. PA 160: Positive x = North,
Negative x = South. PA75 and 75*: Positive x = West, Negative x
= East.}
\end{figure}
\section{Results}
In order to gain a preliminary impression of the variations of the
stellar populations, in Figure 2 we plot the flux ratio:
\begin{equation}
R_B=\frac{\int_{4000}^{4080} F_\lambda d\lambda}{\int_{3620}^{3700}
F_\lambda d\lambda}
\end{equation}
as a function of position along PA 160. This ratio, which is designed
to provide a sensitive measure of the Balmer break, shows remarkably
little variation across the galaxy (typical variation $\pm$ 10\%) except
in the region of the nucleus (within $\pm$ 5'' of the center) which is
subject to enhanced reddening.

To further investigate the stellar populations, 20 apertures were
extracted along the three slits positions with the aim of adequately
sampling the different regions observed in both the 2-D spectra and
the available HST images. These apertures are labelled in Figure
1. Three larger apertures, including all the small apertures for each
slit PA, were also extracted, the latter are labelled as AP$_{TOTAL}$
in Fig 3 and Table 2 and 3. The apertures extracted for the three slit
positions are shown superimosed on spatial profiles of the 2-D spectra
for the wavelength range 4400 -- 4600\AA~ in Figure 3. The apertures
were chosen to be wide enough to have a sufficiently high S/N ratio
for further analysis: S/N $>$ 15. We would like to emphasize that the
S/N is considerably higher than this for most of the apertures
extracted from the different regions of the galaxy. A sample of
extracted spectra is shown in Figure 4. The strong Balmer lines and
Balmer break present across the full measureable extent of the galaxy
(65 arcsec) indicate that young and/or intermediate age stellar
populations are present in all spatial apertures. The overall SEDs
also show signs of enhanced reddening in the nuclear regions
(apertures D and E, PA 160).

In order to obtain a more detailed information about the stellar
populations we have modelled the spectra extracted from the apertures
using the evolutionary synthesis models of \cite{Gonzalez-Delgado05}
and \cite{Bruzual03}. These models incorporate high and intermediate
spectral resolution stellar libraries. Such resolutions are required
to isolate the nebular and the stellar absorption contributions in the
spectra of this type of galaxies. To generate these templates we have
assumed a \cite{Salpeter55} initial mass function, instantaneous
starburst and solar metallicity \cite[see][for a justification of
these assumption]{Tadhunter05}. We then reddened the synthetic spectra
using both the \cite{Calzetti00} reddening law, appropriate for
starburst galaxies, and the \cite{Seaton79} reddening law,
representing the Galactic extinction case. Overall, there is good
consistency between results obtained with the two reddening laws,
demonstrating that the main results are not sensitive to the details
of the reddening law assumed, at optical wavelengths.

To determine the detailed properties of the young stellar populations
we have to fit the whole observed spectrum. We have used two different
approaches to perform the fit and determine the best models
results. They involve STARLIGHT\footnote{www.starlight.ufsc.br}
\citep{Cid-Fernandes05} and CONFIT
\citep{Robinson00,Tadhunter05,Rodriguez-Zaurin07,Holt07} codes. The
former allows up to 11 stellar components but with the same extinction
for all components. On the other hand, the CONFIT approach assumes
only two stellar population components of different age, but allows
these components to have different reddening. More details of the two
techniques are given in sections 3.1 and 3.2.

Throughout this paper, we define young stellar populations (YSPs) as
stellar components with ages t$_{YSP}$ $\leq$ 0.1 Gyr,
intermediate-age stellar populations (ISPs) as stellar components with
ages in the range of 0.1 $<$ t$_{ISP}$ $\leq$ 2 Gyr and old stellar
populations (OSP) as a component with age 12.5 Gyr. Note that we have
not used stellar populations with ages in the range of 2 Gyr $<$
t$_{SP}$ $<$ 12.5 Gyr. Generally, for models that include young and
intermediate/old stellar populations, which is the case for both
CONFIT (2 stellar components plus a power law in some cases) and
STARLIGHT (11 stellar components), it is not possible to distinguish
between ages within this range, and therefore we decided not to use
such stellar populations during the modelling analysis described
below.

\subsection {Results from STARLIGHT}
STARLIGHT fits simultaneously the continuum points and the absorption
lines (high order Balmer lines, He lines, CaII~H, CaII~K,
G-band...). The program reads an input file that labels the spectral
windows that contain emission lines, and these spectral ranges are
excluded of the fitting. In these models, we have only used the SSP
from the evolutionary models\footnote{Starlight can also add a power
law to the SSP to fit the continuum, but these models do not consider
any contribution by a power law}. The intrinsic extinction is modelled
by a foreground dust screen, and parametrized by the V-band extinction,
A$_{\rm V}$. The output is the population vector that represents the
fractional contribution of each SSP of a given age and metallicity to
the model flux at a given wavelength. The fit is carried out with a
simulated annealing plus Metropolis scheme \citep{Cid-Fernandes05},
which searches for the minimum chi-squared between the observations
and the combined models.

For this work we used 11 SSP, corresponding to ages 4, 5, 10, 25, 40,
100, 280, 500, 890 Myr, and 1.25 and 14 Gyr from models by
\cite{Gonzalez-Delgado05}. We use solar metallicity, and the Calzetti
et al. (2000) extinction law. The results obtained are shown in Table
2 and are summarized as follows:
\begin{itemize}
\item OSP contribute very little to the optical continuum. Only a few
percent of the flux at 4020\AA~is due to stellar populations older
than 2 Gyr.
\item ISPs with ages in the range of 0.5 Gyr $<$ t$_{ISP}$ $\leq$ 0.9
Gyr account for $\gsim$ 70\% of the optical continuum in all apertures
apart from apertures D and E in PA 160, sampling the nucleus of the
galaxy.
\item The contribution of YSP ($\leq$ 100 Myr) is significant ($\gsim$
30\%) in the nuclear regions.
\item While low intrinsic reddening is found in the extended regions
(0.0 $\leq$ $E(B - V)$ $\leq$ 0.4 ), large intrinsic reddening, $E(B -
V)$ $\sim$ 0.7 -- 0.8 is required towards the centre of the galaxy.
\item The modelling results found for the three large apertures are
consistent with those of the smaller ones and with each other, as
shown in the table.
\end{itemize}
Similar conclusions are obtained if we use \cite{Bruzual03} synthetic
SEDs. 

A drawback of the STARLIGHT code is that all stellar population
components have the same reddening. Therefore, we have also used the
CONFIT code, allowing the two stellar components used in this approach
to have different reddenings.

\subsection {Results from CONFIT}
The CONFIT approach consists of a direct fit of the overall continuum
shape of the extracted spectra using on a minimum $\chi^2$ technique
\citep{Tadhunter05,Rodriguez-Zaurin07}. CONFIT is based on a
``simplest model'' approach, i.e. we fit the minimum number of stellar
components required to adequately model the data. The main advantage
of this approach is that it allows a wide range of reddening values
for each of the components to be explored. For this work we have used
different combinations of two stellar components (ISP +YSP) with a
range of reddening for each component. For each spectroscopic aperture
the continuum flux was measured in several wavelength bins ($\sim$70)
chosen to be as evenly distributed in wavelength as possible, and to
avoid strong emission lines and atmospheric absorption
features\footnote{Note that it is a further difference with the
STARLIGHT code which fits every pixel apart from regions (e.g. those
containing emission lines or atmospheric emission features) that are
specifically masked out.}. The model fit results are quantified in
terms of the percentage contribution of the different stellar
components in a normalising bin defined as 4700 -- 4800~\AA. Due to
the uncertainties associated with the flux calibration, as well as the
synthetic templates themselves, it is not correct to constrain the
total flux in the normalising bin to be exactly the integrated flux of
the observed spectrum in the bin. Therefore the CONFIT codes allows
the model flux in the normalising bin to vary up to 125\% of the
measured flux in that bin.

Potentially adequate fits to the SEDs (reduced $\chi^2 \lsim 1.0$) can
be found for combinations of small contributions of highly reddened
YSP plus a very old stellar population, or large contributions of low
reddened ISP with an old stellar component contributing much less to
the overall stellar population in the aperture (i.e. the age/reddening
degeneracy problem). To break this degeneracy we have examined
detailed fits to the spectra, considering as valid those models that
fit the important absorption features of the spectra adequately
(e.g. CaII~K, G-band, higher order Balmer lines: see
\cite{Gonzalez01,Holt03,Rodriguez-Zaurin07}).

Table 3 summarizes the results obtained with the CONFIT code.Adequate
fits for all 23 apertures extracted from the three slit positions are
found with models combining two stellar components, an ISP and a YSP,
with no need of an OSP. Figure 4 shows example fits for apertures
selected with the aim of sampling a range of regions of the
galaxy. Note that apertures Ap D and E for PA 160 sample the central
2.5 kpc diameter region, corresponding to dust lanes in the HST
images. It is clear from the figure that the spectra extracted for
these apertures are highly reddened. As well as the best-fitting
models, detailed fits in the region of the Balmer lines (3700 --
4300\AA) are also shown in Figure 4.

Overall, the results reveal a remarkable uniformity in the estimated
stellar ages, with an ISP (0.5 Gyr $<$ t$_{ISP}$ $\leq$ 0.9 Gyr)
dominanting the optical emission in all apertures, and a YSP ($\leq$
0.1 Gyr) making a relatively small contribution ($\leq$ 40\%) in most
apertures. However note that in the central 2.5 kpc of the galaxy (Ap
D and E, PA 160), adequate fits are only found for combinations with a
significant contribution of a YSP component (22 - 63\% ). In terms of 
reddening, while the results found in the extended regions are
consistent with low intrinsic reddening, one or both stellar
components must be highly reddened in the central 5 kpc of the galaxy,
clearly indicating high concentrations of gas and dust in the nuclear
regions.

At this stage, it is important to add a caveat about the spatial
extent of the YSP component. In some of the extended regions of the
galaxy (Ap E and F, PA 75*; Ap A, F and G PA 75; Ap A, PA 160), good
fits are also found with a minimal contribution of a YSP
template. This is consistent with the STARLIGHT results, where the ISP
contribution found for these apertures is $\gsim$ 90 \% of the total
flux. On the other hand, a YSP component is always required to model
the data for the apertures sampling the center of the
galaxy. Therefore, while the presence of a YSP in the extended regions
remains uncertain, the presence of such a component in the galactic
center is essential. This result is consistent with the age of 10-100
Myr \citep{Mundell01} estimated for the duration of the starburst
associated with the galactic superwind \citep{Heckman96} in the
central region of the galaxy, and also with the merger simulations
predictions \citep{Mihos96,Barnes96,Springel05} for starbursts
characteristic of the final stages of major gas-rich mergers.

Again, the results for the three large apertures show remarkable
consistency, both with each other, and with the smaller aperture
results presented in the tables.

Overall, the results obtained with CONFIT are broadly consistent with
those of STARLIGHT approach, and clearly reinforce the idea that an ISP
dominates the optical emission in Arp220.
\subsection {The Hidden Starburst}
Our analysis has so far concentrated on the optically visible star
formation. It is interesting to consider the extent to which this
represents the full extent of star formation in Arp220.

Previous studies based on radio observations \citep{Smith98} or the
galactic superwind \citep{Heckman96,Mundell01}, have found star
formation rates (SFR) up to 100 M$_{\odot}$ yr$^{-1}$ for Arp 220. An
independent estimate of the SFR can be obtained using the published
flux of the Br$\alpha$ line, F(Br$\alpha$) = 1.8 $\times$ 10$^{-13}$
erg s$^{-1}$ cm$^{-2}$ \citep{Genzel98}. Using the IR extinction laws
of \cite{Indebetouw05} and \cite{Rieke85} along with the estimated
visual extinction for the nuclear region of Arp 220 (A$_v$ $\sim$ 45
mag, \cite{Sturm96,Genzel98}), we derive a Br$\alpha$ luminosity of
L(Br$\alpha$) = 9.2 $\times$ 10$^{41}$ erg s$^{-1}$. Asumming Case B
recombination theory we then derive an H$\alpha$ luminosity of
L(H$\alpha$) = 3.3 $\times$ 10$^{43}$ erg s$^{-1}$. It is now
straightforward to estimate the SFR in the galaxy using the
relationship between H$\alpha$ luminosity and SFR given by
\cite{Kennicutt98}. We find a SFR $\sim$ 260 M$_{\odot}$ yr$^{-1}$,
enough to power the entire mid- to far-IR luminosity observed in
Arp220. Moreover, the absence of mid-IR, high-ionization emission
lines ([NeV]$\lambda$14.322$\mu$m, OIV]$\lambda$25.890$\mu$m. Genzel
et al., 1998) suggests no hidden AGN actvity, reinforcing the idea
that starburst activity powers the ULIRG Arp220. In such a case, it is
also possible to measure the SFR of the galaxy using the IR-luminosity
\citep{Kennicutt98}.  For a value of L$_{IR}$ = 1.3 $\times$ 10$^{12}$
L$_{\odot}$ for Arp 220 \citep{Soifer87} we obtain a SFR = 224 M$_{\odot}$
yr$^{-1}$ which is remarkably consistent with the value found using
the Br$\alpha$ flux.

However, there is no clear evidence for the presence of ongoing star
formation activity at optical wavelengths, based on HII region-like
optical emission line ratios \citep{Veilleux99}. In order to measure
the bolometric luminosity associated with the visible stellar
populations detected in the optical, we added the spectra extracted
for all the apertures of the region limited by the dashed-line box in
Figure 1. Since our spectroscopic slits used do not cover the entire
extent of the central region, a large aperture chosen to match the
extent of the box was extracted from an HST ACS (F435W) image (HST
proposal 9396, PI C. Wilson). We then scaled the combined spectrum to
match the flux measured from the aperture, in order to account for
possible flux losses. Assumming an ISP + YSP combination consistent
with the modelling results of section 3.2\footnote{ISP of 700 Myr
($E(B - V) = 0.2$) + a YSP of 5 Myr ($E(B - V) = 0.5$) making a
contribution of 80\% and 20\% of the total flux respectively} and
correcting for reddening effects, we determined the bolometric
luminosity of all the components in the scaled spectrum and added
them, obtaining a total bolometric luminosity L$_{bol} \sim$ 1.6
$\times$ 10$^{11}$L$_{\odot}$. This is an order of magnitude less than
the mid- to far-IR luminosity of the source. Therefore, especially
considering that not all the optical emission is absorbed by dust, it
is clear that the stellar populations detected at optical wavelengths
cannot power the far-IR luminosity of the source; most of the ongoing
star formation in the nuclear regions is hidden by dust.
\subsection {Summary of the key results}
To summarize tha main results, the modelling of our spectra provide
evidence for three stellar populations in Arp220:
\begin{itemize}
\item {\bf Dominant ISP.} We have found an intermediate-age stellar
  population, with ages 0.3 Gyr $<$ t$_{ISP}$ $\leq$ 0.9 Gyr, that
  dominates the optical spectrum, extended across 65 arcsec ($\sim$ 24
  kpc), and covering the measurable extent of the galaxy.
\item {\bf YSP component.} We have found a young stellar component,
  with ages t$_{YSP}$ $\leq$ 100 Myr, with varying contributions
  across the galaxy, but particularly significant in the nuclear
  regions, where it is highly reddened.
\item {\bf Hidden Starburst.} Our results reinforce the idea that
  there is prodigious ongoing star formation activity that is hidden
  at optical wavelengths, likely to be related to the radio supernovae
  detected in the center of the galaxy \citep{Smith98}, as well as the
  powerful mid to far-IR emission from this object.
\end{itemize}
\begin{table*}
\centering
\begin{tabular}{@{}llccccc@{}}
\hline   &&& STARLIGHT &&&\\ 
\hline   
         &  &Width of   & $E(B -V)$    & \%YSP      &\%ISP      &\% OSP \\
         &  &the Aperture  &              &            &           &\\               
         &  &(Arcsec)   &              &            &           &\\
         (1) & (2) & (3) & (4) & (5) & (6) & (7) \\
\hline
PA 160    &Ap A &-12.6 to -8.2& 0.0 & 0  & 112  & 0\\
          &Ap B &-8.2 to -5.4 & 0.1 & 0  & 109  & 0\\
          &Ap C &-5.4 to -1.8 & 0.4 & 10 & 96   & 0\\
          &Ap D &-1.8 to 1.8  & 0.8 & 58 & 34   & 9\\
          &Ap E &1.8 to 4.6   & 0.7 & 35 & 63   & 5\\
          &Ap F &4.6 to 8.6   & 0.4 & 30  & 71  & 2\\
          &Ap G &8.6 to 22.6  & 0.2 & 13  & 91  & 0\\
          &AP$_{TOTAL}$ &-24.4 to 28.4 & 0 & 25 & 80 & 1\\ 
\hline   
PA 75     &Ap A &-26.6 to -8.6& 0.3 & 10  & 96   & 0 \\
          &Ap B &-8.6 to -4.2 & 0.5 & 33  & 68  & 2 \\
          &Ap C &-4.2 to -1.4 & 0.5 & 25 & 77   & 2 \\
          &Ap D &-1.4 to 1.4  & 0.4 & 16 & 88   & 0 \\
          &Ap E &1.4 to 5.0   & 0.4 & 21 & 84   & 1\\
          &Ap F &5.0 to 10.2  & 0.4 & 15  & 87  & 0 \\
          &Ap G &10.2 to 34.6 & 0.3 & 21  & 91  & 0 \\
          &AP$_{TOTAL}$ &-29.4 to 35.8 & 0.4 & 18 & 87 & 0\\    
\hline 
PA75*     &Ap A &-19.2 to -12.0 & 0.2 & 16 & 89  & 0\\
          &Ap B &-12.0 to 5.2   & 0.2 & 9 &  96  & 0\\
          &Ap C &-5.2 to -2.8   & 0.2 & 15 & 89  & 0\\
          &Ap D &-2.8 to 2.8    & 0.2 & 16 & 90  & 0\\
          &Ap E &2.8 to 17.6    & 0.2 & 17 & 91  & 0\\
          &Ap F &17.6 to 38.4   & 0.2 & 14 & 93  & 0\\
          &AP$_{TOTAL}$ &-23.2 to 45.6 & 0 & 14  & 92 & 0\\
\hline
\hline
\end{tabular}
\caption{Modelling results obtained using the STARLIGHT code with the
Gonz\'alez Delgado et al. (2005) synthetic SEDs. Col (1): Slit PA. Col
(2): Aperture label. Apertures D and E from PA 160 sample the nuclear
regions of the galaxy Col (3): Width of the aperture (the centroid of
Ap D is chosen to be the reference point for all the slit
positions). Col (4): The intrinsic reddening value obtained for the
best fitting models assuming the same extinction for all the stellar
components. Cols (5), (6) and (7): The flux percentage at 4020\AA~of
each population, as defined in Section 3.}
\end{table*}
\begin{table*}
\centering
\begin{tabular}{@{}llcccccc@{}}
\hline                      & && CONFIT &&&\\   

\hline    
         &   & Width of    &Age of    &E(B - V)   & Age of &E(B -V) &\%YSP\\ 
         &   & the Apert   &ISP       &           &YSP              & &of total\\
         &   & (arcsec)   &(Gyr)      &           &(Gyr)   &        &flux \\  
          (1) & (2) & (3) & (4) & (5) & (6) & (7) & (8) \\
\hline  
PA 160  &Ap A &-12.6 to -8.2 & 0.6 - 0.7 & 0.0 - 0.2 & $<$ 0.1 & 0.0 - 0.2 & $<$ 5\\     
        &Ap B & -8.2 to -5.4 & 0.5 - 0.7 & 0.0 - 0.2 &  $<$ 0.1 & 0.0 - 0.3 & $<$ 20\\
        &Ap C &  -5.4 to -1.8 & 0.5 - 0.7 & 0.3 - 0.5 &  $<$ 0.1 & 0.0 - 1.0 & 5 - 15 \\   
        &Ap D & -1.8 to 1.8 & 0.5 - 0.9 & 0.0 - 1.0 &  $<$ 0.08 & 0.9 - 1.8 & 30 - 63\\
        &Ap E & 1.8 to 4.6 & 0.5 - 0.9 & 0.2 - 0.6 &  $<$ 0.06 & 0.9 - 1.6 & 22 - 47\\
        &Ap F & 4.6 to 8.6 & 0.5 - 0.7 & 0.2 - 0.6 & $<$ 0.08 & 0.0 - 1.5 &  5 - 36\\
        &Ap G & 8.6 to 22.6 & 0.5 - 0.7 & 0.0 - 0.3 &  $<$ 0.07 & 0.0 - 0.7 & 5 - 20 \\  
        &AP$_{TOTAL}$ & -24.4 to 28.4 & 0.5 - 0.7 & 0.2 - 0.6 & $<$ 0.1 & 0.0 - 1.5 &  5 - 35 \\ 
\hline 
PA 75   &Ap A & -26.6 to -8.6 & 0.5 - 0.7 & 0.2 - 0.5 & $<$ 0.1 & 0.0 - 2.0 &    $<$ 20 \\ 
        &Ap B & -8.6 to -4.2 & 0.5 - 0.7 & 0.2 - 0.6 & $<$ 0.1 & 0.0 - 1.0 & 5 - 25 \\   
        &Ap C & -4.2 to -1.4 & 0.5 - 0.7 & 0.2 - 0.6 & $<$ 0.05 & 0.0 - 1.2&    5 - 30 \\ 
        &Ap D & -1.4 to 1.4 & 0.5 - 0.7 & 0.2 - 0.6 & $<$ 0.06 & 0.0 - 1.5&    5 - 25 \\ 
        &Ap E & 1.4 to 5.0 & 0.5 - 0.7 & 0.2 - 0.5 & $<$ 0.05 & 0.0 - 2.0&    $<$ 30 \\ 
        &Ap F & 5.0 to 10.2 & 0.6 - 0.9 & 0.1 - 0.4 & $<$ 0.1 & 0.0 - 1.2& $<$ 25 \\
        &Ap G & 10.2 to 34.6 & 0.7 - 0.9 & 0.0 - 0.3 & $<$ 0.1 & 0.0 - 0.8& $<$ 25 \\ 
        &AP$_{TOTAL}$ & -29.4 to 35.8 & 0.5 - 0.7 & 0.2 - 0.5 & $<$ 0.1 & 0.0 - 1.2& 5 - 25 \\
\hline 
PA75*   &Ap A & -19.2 to -12.0 & 0.5 - 0.7 & 0.0 - 0.3 & $<$ 0.06 & 0.0 - 0.6 & 10 - 35 \\  
        &Ap B & -12.0 to 5.2 & 0.5 - 0.7 & 0.0 - 0.3 & $<$ 0.1 & 0.0 - 1.5  &   5 - 25  \\  
        &Ap C & -5.2 to -2.8 & 0.5 - 0.7 & 0.0 - 0.3 & $<$ 0.1 & 0.0 - 1.5  &   $<$ 40  \\
        &Ap D & -2.8 to 2.8 & 0.5 - 0.8 & 0.0 - 0.3 & 0.06 & 0.0 - 1.5 &   $<$ 40  \\  
        &Ap E & 2.8 to 17.6 &0.6 - 0.9 & 0.0 - 0.3 & $<$ 0.06 & 0.0 - 1.0& $<$ 25  \\ 
        &Ap F & 17.6 to 38.4 & 0.6 - 0.9 & 0.0 - 0.3 & $<$ 0.1 & 0.0 - 2.0& $<$ 15  \\ 
        &AP$_{TOTAL}$ & -23.2 to 45.6 & 0.5 - 0.8 & 0.0 - 0.3 & $<$ 0.06 & 0.0 - 1.5&   5 - 30  \\
\hline \hline
\end{tabular}
\caption{Modelling results obtained using the CONFIT code with the
Bruzual \& Charlot (2003) synthetic SEDs. Col (1): Slit PA. Col (2):
Aperture label. Apertures D and E from PA 160 sample the nuclear
regions of the galaxy Col (3): Width of the aperture (the centroid of
Ap D is chosen to be the reference point for all the slit
positions). Col (4): Range of ages for the ISP component. Col (5):
Range of intrinsic $E(B - V)$ values for the ISP component. Col (6):
Upper limits for the ages of the YSP component. Col (7): Same as
column 5 for the YSP component. col (8): Upper limits for the
contribution in flux of the YSP to the model in the normalising bin
(4700\AA~ --- 4800\AA), for those cases for which an ISP can model the
data with a negligible contribution of a young component.}
\label{Modelling results}
\end{table*}
\begin{figure*}
\vspace*{0cm}\psfig{file=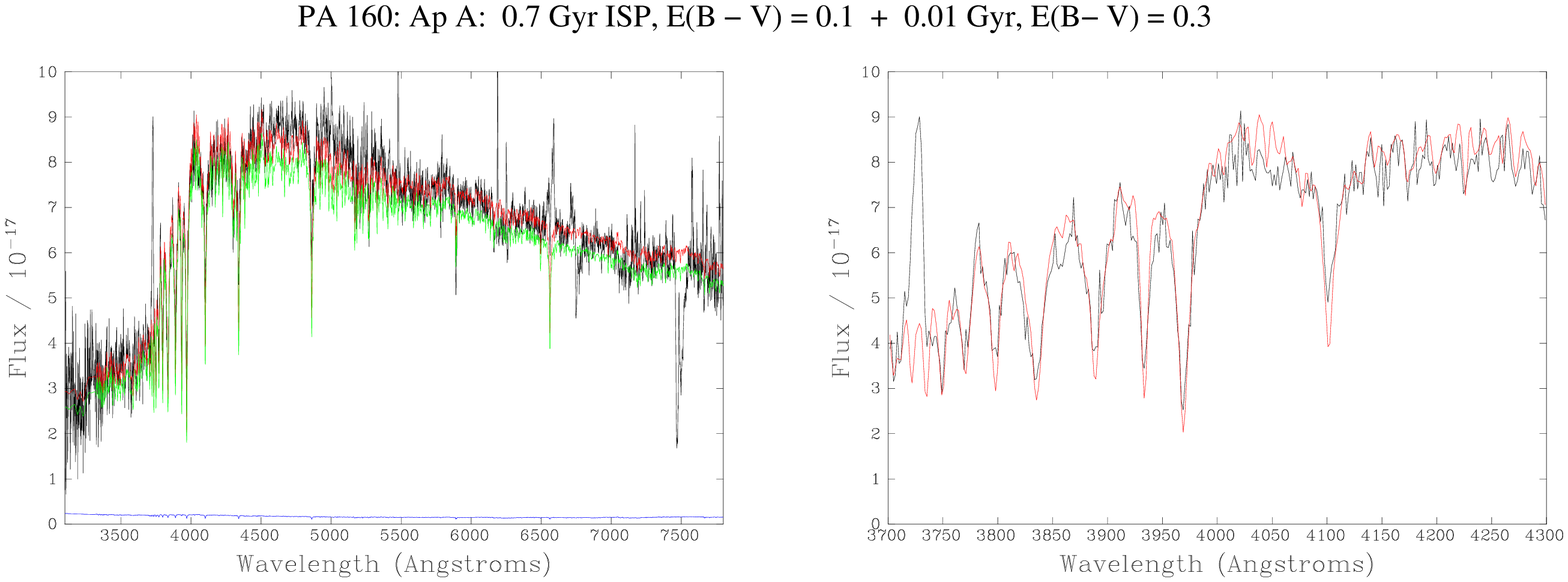,width=15.0cm}\\
\vspace*{0.5cm}\psfig{file=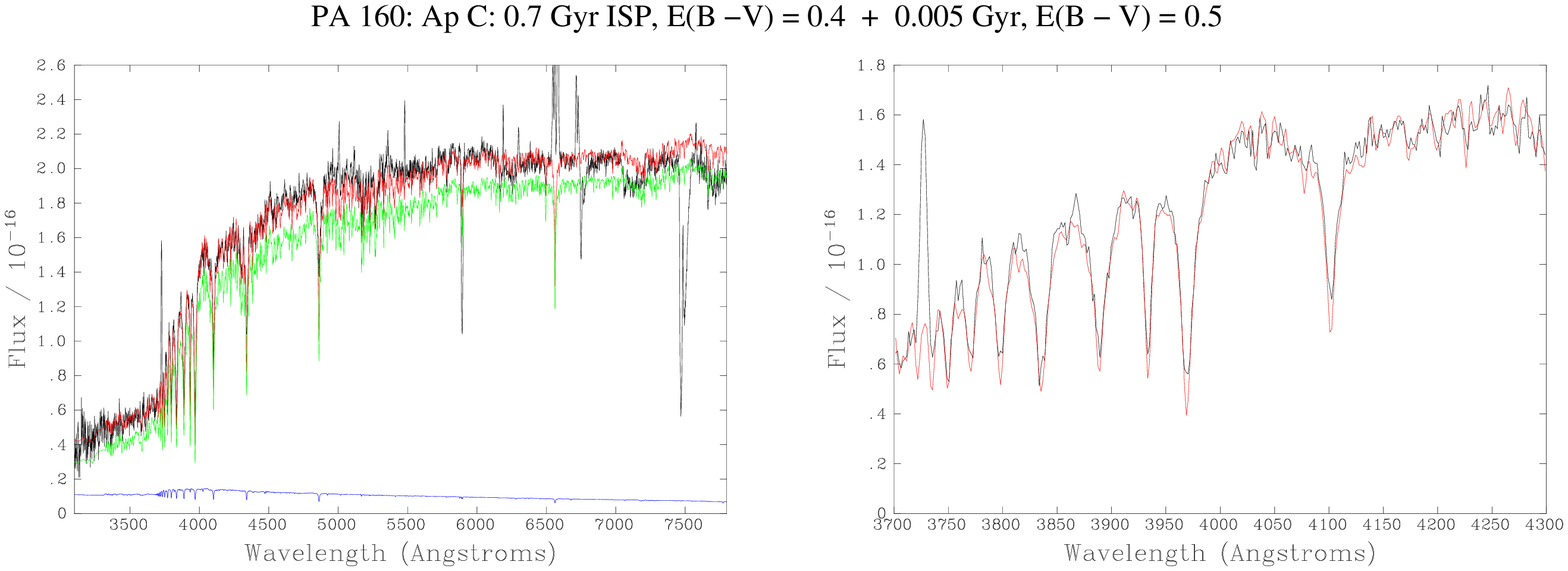,width=15.0cm}\\
\vspace*{0.5cm}\psfig{file=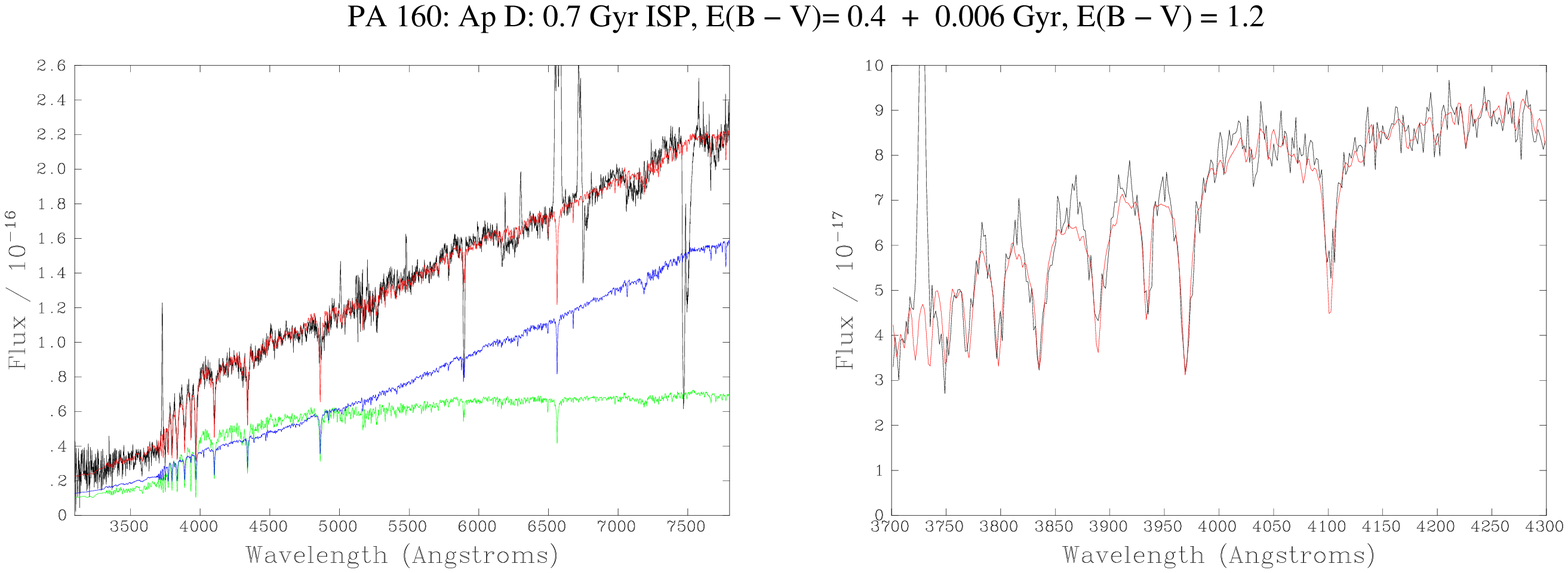,width=15.0cm}\\
\vspace*{0.5cm}\psfig{file=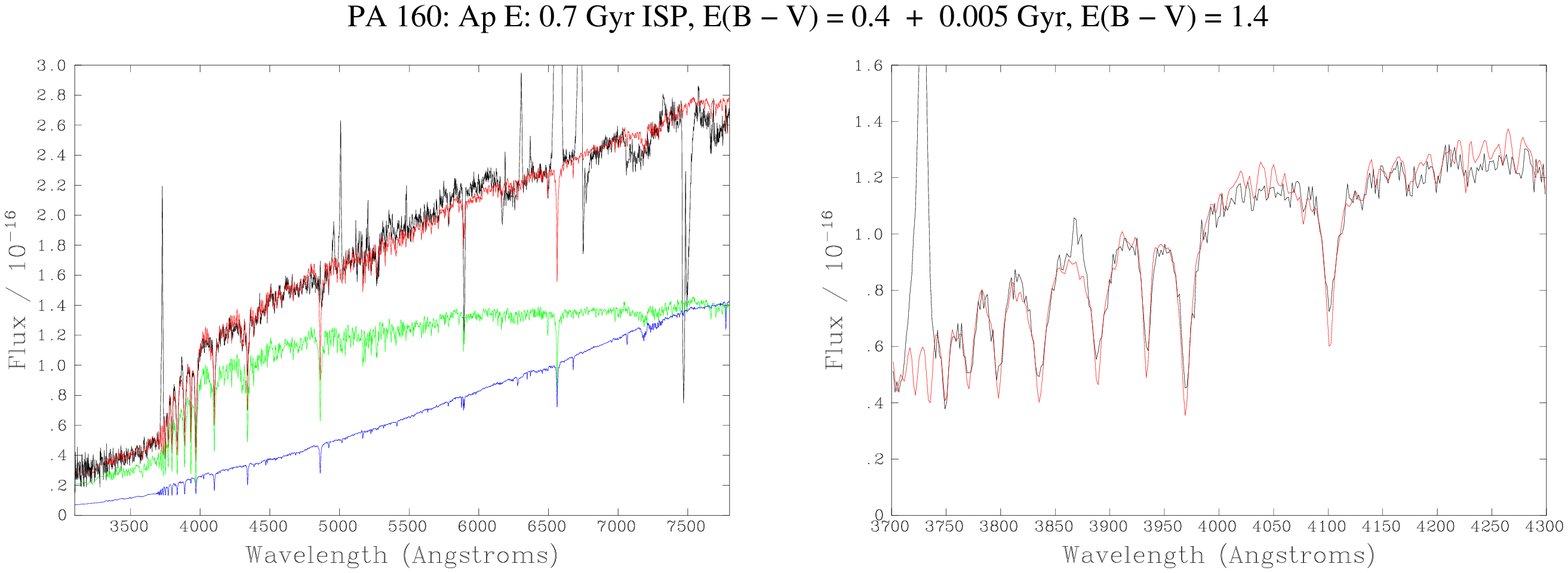,width=15.0cm}\\
\caption[]{}
\label{fig:SED}
\end{figure*}
\addtocounter{figure}{-1}
\begin{figure*}
\vspace*{0.0cm}\psfig{file=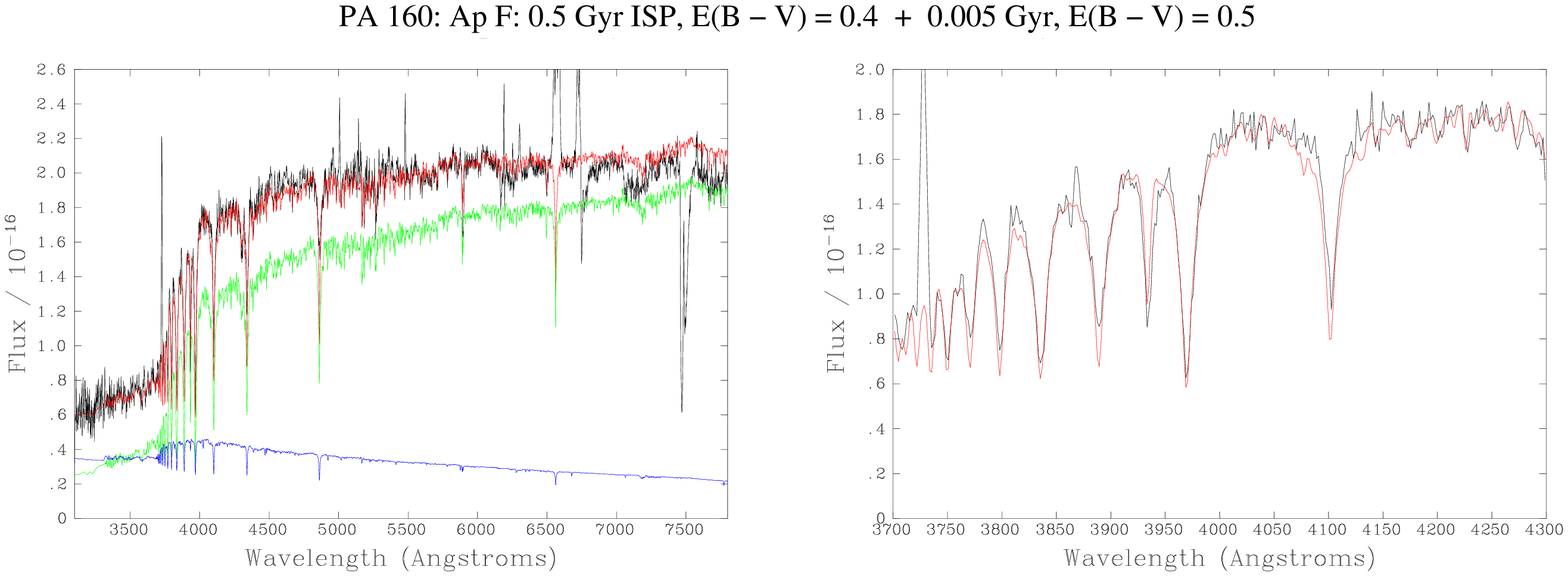,width=15.0cm}\\
\vspace*{0.5cm}\psfig{file=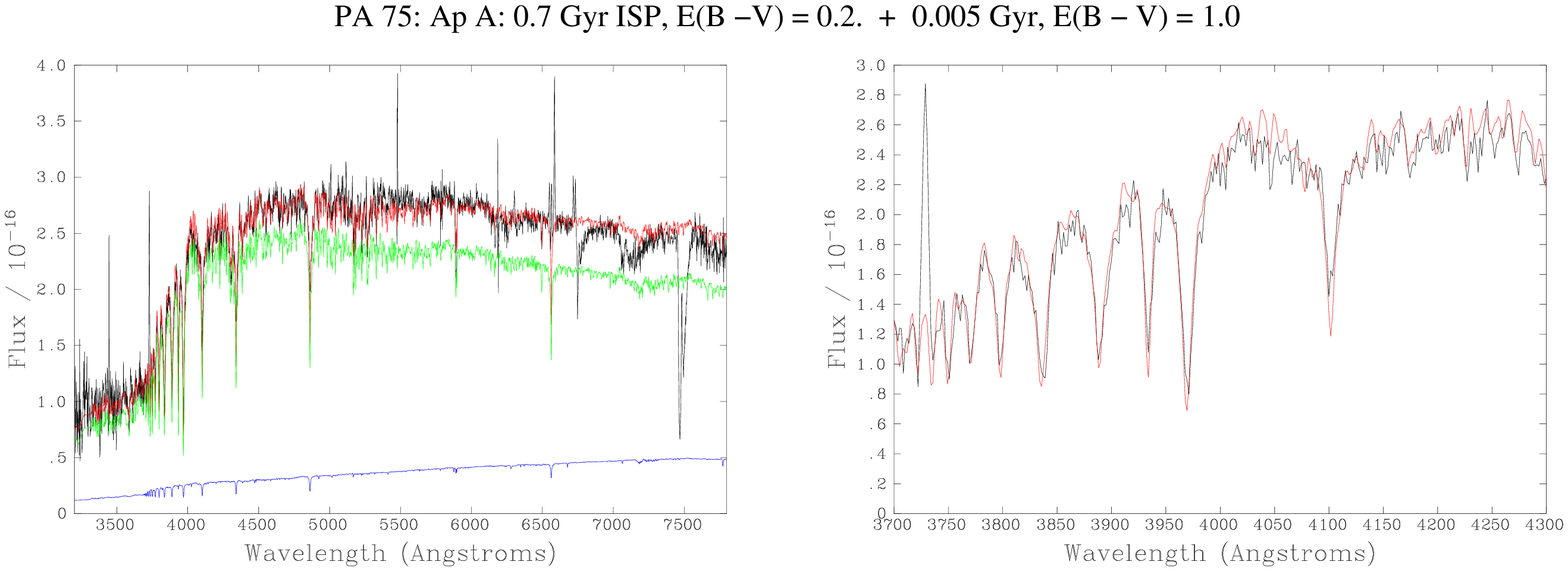,width=15.0cm}\\
\vspace*{0.5cm}\psfig{file=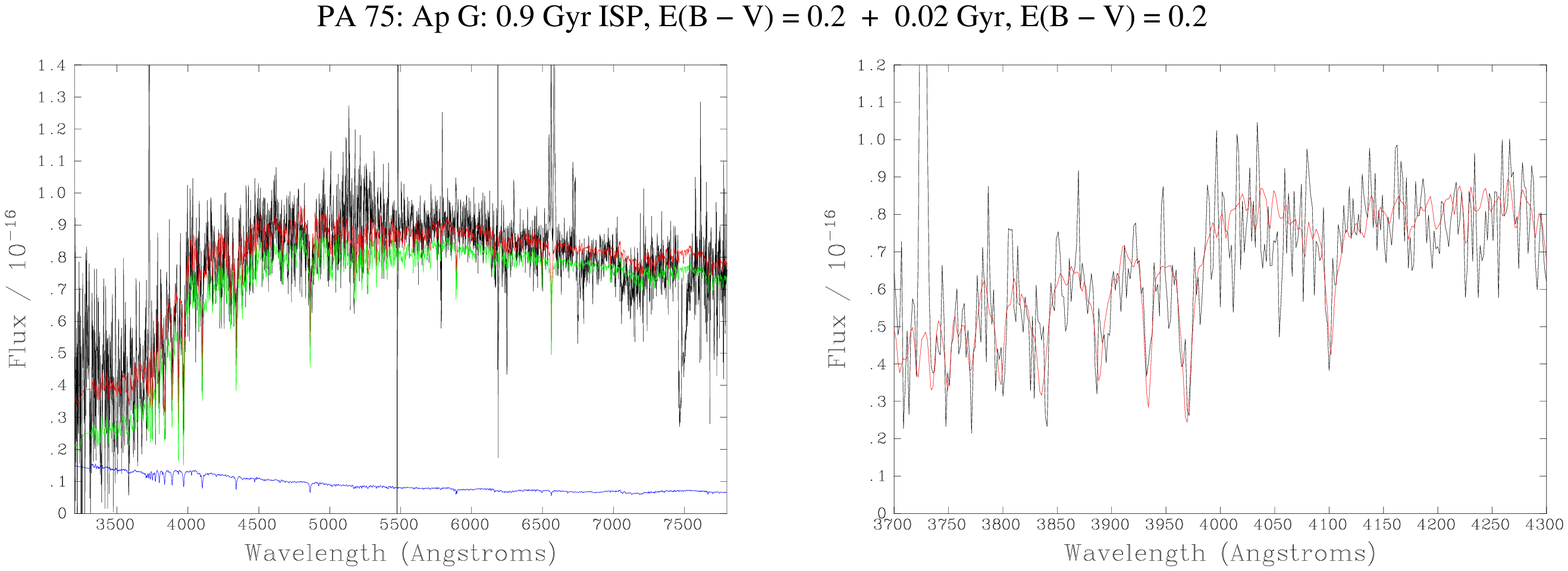,width=15.0cm}\\
\vspace*{0.5cm}\psfig{file=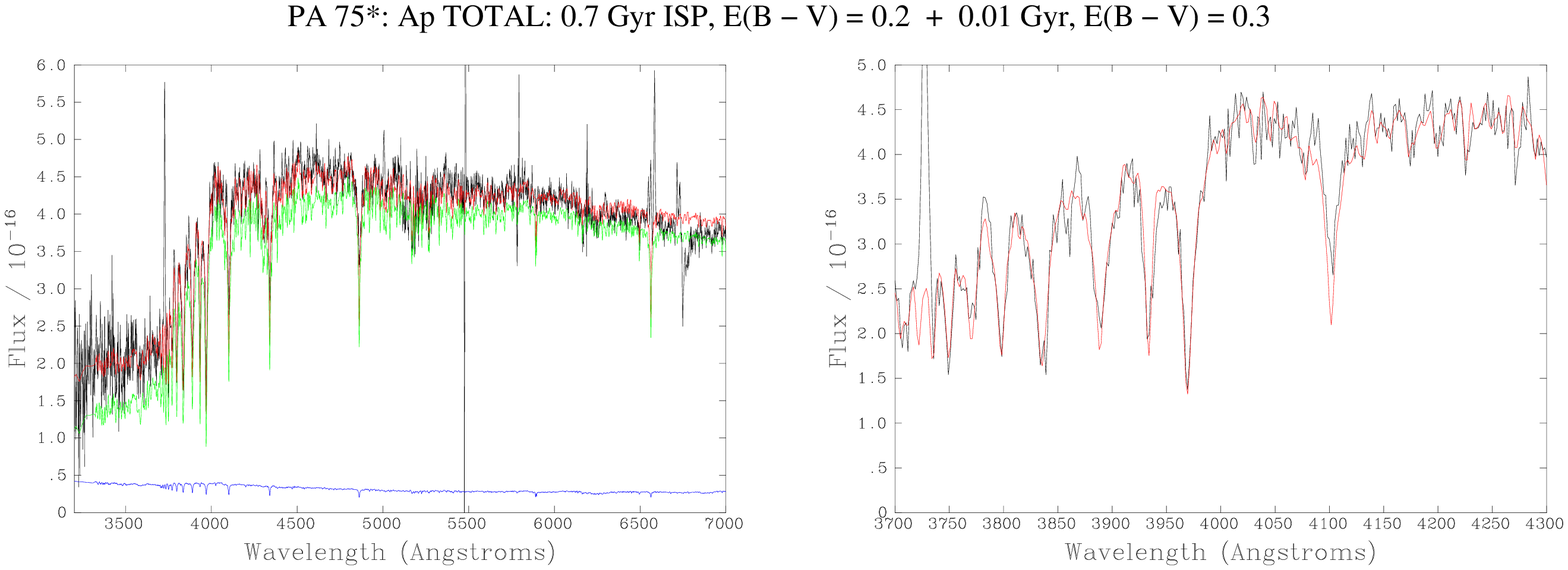,width=15.0cm}\\
\caption[]{Detailed fits to the spectra extracted for some of the
apertures sampling the different regions of the galaxy (see Figure 1
for the location of the extraction apertures). The green and blue
lines are the ISP and YSP components respectively, while the red lines
are the sum of both. It is clear from the figure that a variety of
reddenings is required to fit the data for the different
apertures. Also shown in the figure are detailed fits of the
wavelength range $3700\AA~ - 4300\AA$ (plots on left). The fluxes are
presented in wavelength units.}
\label{fig:SED}
\end{figure*}

\section{Discussion}
\subsection{Comparison with previous results}
\cite{Wilson06} found evidence for a centrally located YSP ($\leq$ 10
Myr) and an intermediate-age ($\sim$ 300 Myr) stellar population
spread towards the north of the galaxy in their photometric study of
the cluster populations in Arp 220. These results are broadly
consistent with the results presented in this paper. In addition, we
find evidence for ISPs at all locations in the galaxy.

There have been relatively few attempts in the past to study the
stellar populations in the diffuse light of ULIRGs.
\cite{Canalizo00a,Canalizo00b,Canalizo01} found stellar ages ranging
from currently ongoing starburst activity to poststarburst ages of
$\lsim$300 Myr in their sample of ``transition QSOs'' --- objects that
may represent a transitionary stage between ULIRGs and QSO. The ages
found for the dominant stellar populations in Arp220 are generally
older than 300 Myr, although \cite{Canalizo00a,Canalizo00b,Canalizo01}
used only one stellar component for their modelling.

More recently, \cite{Rodriguez-Zaurin07} applied the same techniques
used here to the ULIRG/radio galaxy PKS1345+12. Overall, the results
found in their paper are consistent with the results found here for
Arp 220, with a mixture of YSP and ISP components. However, precise
age determination for the ISP component is more difficult in
PKS1345+12 because of the presence of a strong OSP component.

We are currently analyzing the optical spectra of a sample of 40
ULIRGs (Rodriguez et al., 2007, in prep). We find that the modelling
results are consistent with a combination of YSP + ISP components for
the majority of the objects in our sample. However, we find that the
YSPs found in these galaxies make a larger contribution than in the
case of Arp220, and sometimes dominate the optical spectra. Finally,
similar results in terms of the mix of stellar populations have
recently been obtained in detail spectroscopic studies of LIRGs
\citep{Gonzalez-Delgado07}.
\subsection{The origin of the ISPs}
In order to use studies of the stellar populations to investigate star
forming histories in major galaxy merges it is important to understand
the extent to which the stellar populations have been formed during
the merger. A potential issue is contamination by the young stellar
populations present in the galaxy disks prior to the star of the
merger. As a first approach to investigate the origin of the ISPs we
used the CONFIT code and combined unreddened Sa, Sb and Sc galaxy
template spectra \citep{Kinney96} with a stellar population with ages
in the range 0.001 $\leq$ age $\leq$ 5 Gyr and reddenings 0.0 $\leq
E(B - V)\leq$ 2.0.

No good fits were found for combinations of Sa, Sb or Sc galaxy
templates without a significant additional contribution of a YSP or an
ISP component. However, the \cite{Kinney96} templates represent an
average for several galaxies of the same morphological type, and each
type encompasses a range of individual galaxy spectra, some of which
deviate substantially from the averaged SED. Therefore, although the
modelling results obtained using this combination suggest that the
stellar populations detected in the optical for the ULIRG Arp220 have
been formed during the merger event rather than captured, we cannot
entirely rule out the idea that the ISPs detected in the galaxy
represent the disrupted disks of one or more of the merging galaxies.

Another way to address this issue is to compare the mass of YSPs in
late type spiral galaxies, the likely progenitor galaxies, with the
values found here for the ISPs in Arp 220. The reason to focus only on
YSPs in the parent galaxies is that, if the star formation in the
disks is truncated during the merger, these are the populations which
may evolve into ISPs detected today in Arp 220. Since our three slit
positions do not cover the full extent of the object, we used the
apparent magnitude of the galaxy in the V band, m$_V = 13.2$
\citep{Taylor05}, and applied the same techniques described in section
3.4 to account for possible slit losses. We obtain total masses in the
range of 2.7$\times$10$^{10}$ -- 4.3$\times$10$^{10}$M$_{_{\odot}}$. for
the ISPs in Arp220. To estimate the stellar mass content of late type
spiral galaxies, we modelled a Sc galaxy template \citep{Kinney96}
combining a 12.5 Gyr OSP with a YSP. We find that we can model the
template with an OSP plus YSPs of age t$_{YSP}$$\leq$ 20 Myr
contributing 70 - 100\% of the total flux in the normalising bin
(4420\AA -- 4500\AA, chosen to be close to the B band wavelength
range). We then assumed the same values for the contribution of each
component to the median B-band monochromatic luminosities of Sc
galaxies \citep{Roberts94}. Since the templates are generated for 1
M$_{_{\odot}}$, by scaling we were able to estimate the total mass and
the mass contribution of each of the components. We estimated a range
of 3.6$\times$10$^{9}$ -- 3.7$\times$10$^{11}$M$_{_{\odot}}$ for the
total stellar mass of the population of typical Sc galaxies in the
large compilation of \cite{Roberts94}, of which only
1.2$\times$10$^{9}$ -- 8.5 $\times$10$^{9}$M$_{_{\odot}}$ is accounted by
the YSP component. Considering only the top 25\% of the most massive
Sc galaxies in the \cite{Roberts94} sample, the latter values increase
by a factor of $\lsim$ 2. Therefore, it is possible that the ISPs
detected in Arp220 represent the captured disks of one or more of the
parent galaxies, but only if these galaxies are among the most massive
late type spirals.

In terms of gas content, \cite{Scoville97} found a mass of $\sim$ 3
$\times$10$^{10}$M$_{_{\odot}}$ for the H$_{2}$ mass of Arp 220. However,
they used a standard Galactic CO-to-H$_{2}$ ratio, which is likely to
overestimate the molecular gass mass M(H$_{2}$) by a factor of three
\citep{Solomon97}. On the other hand, an upper limit for the HI mass
value for Arp220 is 5.0$\times$10$^{9}$M$_{_{\odot}}$
\citep{Mirabel88}. Therefore, the total gas mass in the galaxy,
M(H$_{2}$ + HI), is of order of $\sim$ 1.5 $\times
$10$^{10}$M$_{_{\odot}}$. In comparison, the median value of the HI mass
content of Sc galaxies is 0.8$\times$10$^{10}$M$_{_{\odot}}$
\citep{Roberts94}. Asuming a typical H$_{2}$/HI mass ratio of 0.7
\citep{Young89} for such galaxies, we find a gas mass content of
M(H$_{2}$ + HI) $\sim$ 1.4$\times$10$^{10}$M$_{_{\odot}}$ for a typical
Sc galaxy. Therefore, we conclude that it is possible for a merger of
two typical spiral galaxies to account for the gas content estimated
in Arp220 provided that a large fraction of the gas is not ejected
during the merger \citep{PDiMatteo} or transformed into stars. However
if a substantial proportion of the ISPs are formed in the merger, then
the total amount of gas required is much larger, implying a merger
between two Sc galaxies in the upper 25\% of the HI mass range
\citep{Roberts94}.

\subsection{Comparison with models}
Arp220 is a double nucleus system with a separation of 0.95 arcsec,
corresponding to $\sim$345 pc, that exhibits tidal tails in the
optical. The morphology of the galaxy suggests that it is in the final
stages of the merger event, just before the two nuclei coalesce. In
this section we will compare the results found in this paper with the
prediction of the models for the star formation activity at such a
stage of a merger event.

In general, simulations predict two epochs of starburst activity
\citep{Mihos96,Barnes96,Springel05} in major gas-rich mergers: the
first ocurring just after the first encounter, and the second, more
intense, episode towards the end of the merger, when the nuclei
coalesce. However, both the time lag and the relative intensity of the
peaks of starburst activity during the merger event depend on several
factors: the presence of bulges, feedback effects, gas content and
orbital geometry. For example, the presence of a bulge acts as a
stabilizer of the gas against inflows and formation of bar structures,
allowing stronger starburst activity towards the end of the merger
event \citep{Mihos96,Barnes96}. On the other hand, AGN feedback
effects (e.g. quasar-driven winds) disrupt the gas surrounding the
black hole, acting against the star formation activity
\citep{Springel05}.

In the case of Arp220, we derive star formation rates of $\sim$ 260
M$_{\odot}$ yr$^{-1}$, consistent with a starburst being responsible for
the IR luminosity of the galaxy and there is no evidence for AGN
activity \citep{Genzel98}. Based on the results of the simulations
\citep{Springel05}, highly gas-rich disc galaxies must be involved in
the merger event to account for such high star formation rates. In
this context, the results found in the previous section are consistent
with the model predictions. It is possible that the dominant ISPs
detected in Arp220 are associated with the first enhancement in star
formation activity that occurs in the early stages of the merger,
coinciding with the first encounter between the merging
nuclei. However, because of the potential contamination from stars in
the disrupted disks of the merging galaxies (section 4.2), the
fraction of ISPs formed in the merger is not known with any
accuracy. On the other hand, the YSPs detected in the center of the
galaxy, as well as the ULIRG activity, are likely to be related to the
major enhancement of the activity as the two nuclei coalesce in the
final stages of the merger. It is notable that the ages of the YSP
(t$_{YSP}$ $<$ 0.1 Gyr) are consistent with the 0.1 Gyr timescale of
the final starburst predicted by the merger simulations.

\section{Conclusions}
The results presented in this paper demonstrate the utility of
spectroscopic studies of the diffuse light for investigating the
evolution of the stellar populations in ULIRGs.

We have found evidence for a complex star forming history in the ULIRG
Arp 220, with at least two distinct episodes of starburst
activity. The results are consistent with an intermediate-age stellar
population of age 0.1 $<$ t$_{ISP}$ $\leq$ 0.9 Gyr, dominating the
optical emission throughout the body of the galaxy. We have also
detected young ($\leq$ 100 Myr) stellar populations (YSPs) located in
the central $\sim$ 2.5 kpc that contribute significantly to the
optical emission. This latter population is likely to be related to
the last enhancement of the activity towards the end of the merger
event.

On the other hand, we have estimated that the bolometric luminosity
accounted by the optically visible stellar populations and compared it
with the IR-luminosity of the galaxy. We conclude that we are not
directly detecting the bulk of the starburst activity responsible for
the ULIRG phenomenon in Arp220, which is likely to be related to the
radio supernovae found by \cite{Smith98}, representing another
``extra'' episode of star formation activity apart from the two
suggested by our modelling results. It is clear that our results for
Arp220 are consistent with the emerging evidence for a complex and
multimodal star formation activity in merging systems ( The Antennae:
\cite{Whitmore99}, NGC7252: \cite{Maraston01}, PKS1345+12:
\cite{Rodriguez-Zaurin07}).

In terms of reddening, while the presence of highly reddened stellar
component is unlikely in the extended regions of the galaxy, high
reddening values are required to match the data for the apertures
sampling the nuclear regions, coinciding with dust lanes in the HST
images. Our results show the importance of accounting for reddening
when trying to model stellar populations in star forming galaxies.

Finally, we note that the modelling results for the three large
apertures, representing the integrated light of the galaxy, are
consistent with those of the detailed study based on smaller
apertures. This gives confidence to studies of ULIRGs at medium and
high redshifts, suggesting that age measurements based on relatively
large aperture studies of the diffuse light at such redshifts, are
representative of the stellar populations of the galaxies as a whole.

\section*{Acknowledgments} 
JR acknowledges financial support in the term of a PPARC
studentship. RGD is supported by the Spanish grant AYA2004-02703. The
William Herschel Telescope is operated on the island of La Palma by
the Isaac Newton Group in the Spanish Observatorio del Roque de los
Muchachos of the Instituto de Astrofisica de Canarias. We thank Paul
Crowther for useful discussions and Roberto Cid for STARLIGHT. We also
thank the referee for her useful comments which have helped to improve
the manuscript.

\bibliographystyle{mn2e}
\bibliography{javi}

\end{document}